\documentclass[aps,prb,reprint,superscriptaddress]{revtex4-1}
\usepackage[dvipdfmx]{graphicx} 
\usepackage{url}
\usepackage{dcolumn}
\usepackage{bm,braket}
\usepackage{xcolor}

\begin{document}

\title{Quantum critical dynamics in two-dimensional transverse Ising model}
\author{Chisa Hotta}
\affiliation{Department of Basic Science, University of Tokyo, Komaba, Meguro-ku, Tokyo 153-8902, Japan}
\author{Tempei Yoshida}
\affiliation{Department of Physics, Kyoto Sangyo University, Kamigamo Motoyama, Kyoto 464-8603, Japan}
\author{Kenji Harada}
\affiliation{Graduate School of Informatics, Kyoto University, Kyoto 606-8501, Japan}

\begin{abstract}
In the vicinity of the quantum critical point(QCP), 
thermodynamic properties diverge toward zero temperature governed by universal exponents. 
Although this fact is well known, how it is reflected in quantum dynamics has not been addressed. 
The QCP of the transverse Ising model on a triangular lattice is an ideal platform to test the issue, 
since it has an experimental realization, the dielectrics realized in 
an organic dimer Mott insulator, $\kappa$-ET$_2X$, 
where a quantum electric dipole represents the Ising degrees of freedom. 
We track the Glauber-type dynamics of the model by constructing a kinetic protocol based on the quantum Monte Carlo method. 
The dynamical susceptibility takes the form of the Debye function and shows a significant peak-narrowing 
in approaching a QCP due to the divergence of the relaxation timescale. 
It explains the anomaly of dielectric constants observed in the organic materials, 
indicating that the material is very near the ferroelectric QCP. 
We disclose how the dynamical and other critical exponents develop near QCP beyond the simple field theory. 
\end{abstract}
\maketitle
%
\begin{figure}[t]
  \centering
  \includegraphics[width=90mm]{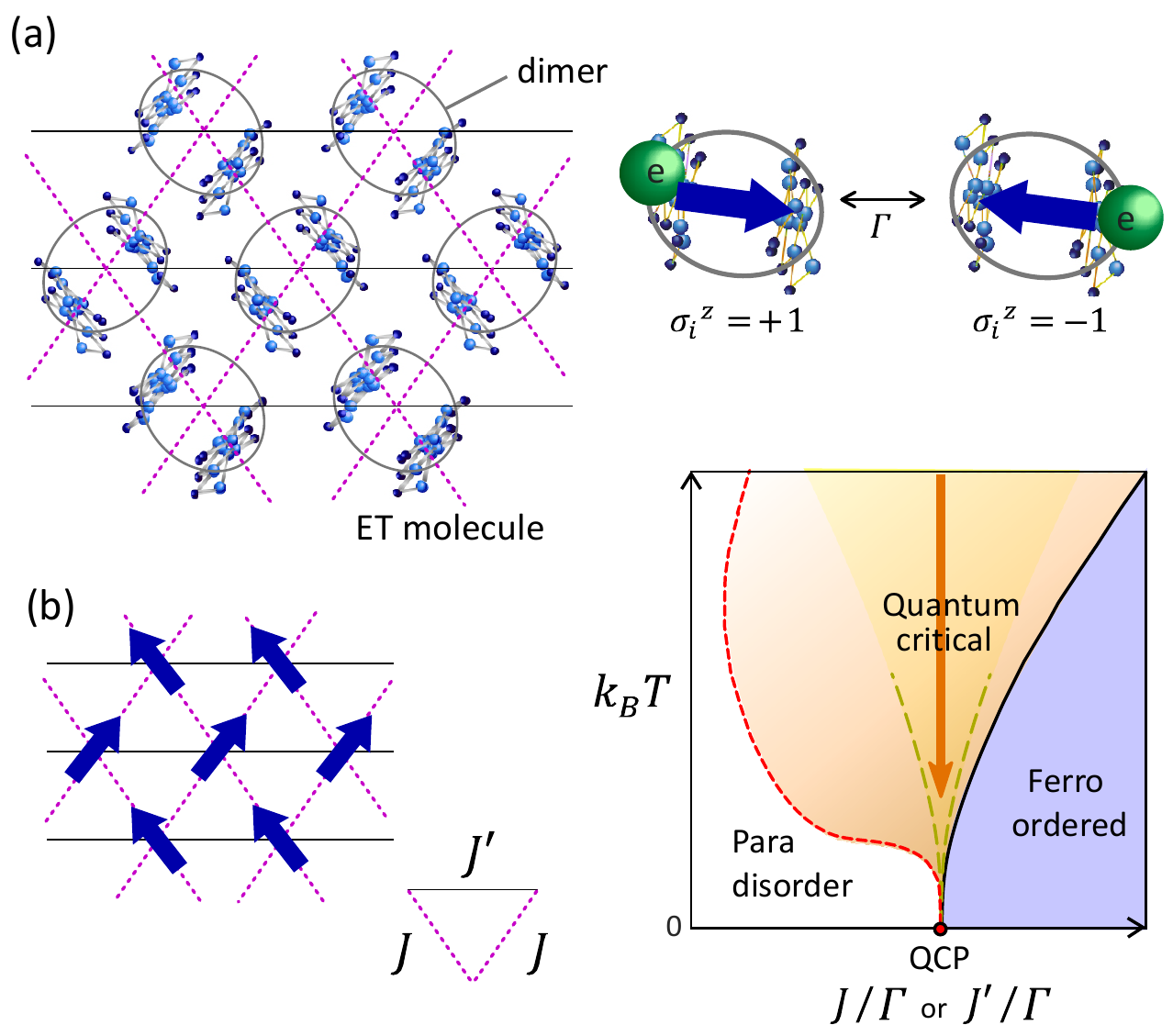}
  \caption{
    (a) Two-dimensional electronic plane of $\kappa$-(ET)$_2X$. 
    The electric dipole (right panel) is defined on each dimer as arrows that could point 
    in two different directions depending on the location of the charges. 
    As the charge hops between two molecules, the dipole 
    fluctuates quantum mechanically by $\Gamma$. 
    (b) Schematic description of the transverse Ising (TRI) model on an anisotropic triangular
    lattice. Arrows indicate the Ising degrees of freedom. 
    Phase diagram (right panel) near QCP for the material parameters of $\kappa$-(ET)$_2X$ 
    extracted from our results(see Fig.~\ref{f3}). 
    The broken yellow line is the phenomenologically discussed crossover line of the critical region, 
    whereas the region marked with a red dot line is the one obtained from our calculation 
    where the static and dynamical properties, $\chi_0$ and $\tau$, take enough large values. }
  \label{f1}
\end{figure}
\section{Introduction}
Criticality is a phenomenon characterized by an algebraically growing fluctuation that spreads throughout the system 
and eventually manifests as a scale invariance of the physical properties\cite{Cardy1996}. 
Thermodynamic properties behave critically as the system approaches the second order phase transition point 
which is detected by the divergence of the specific heat and susceptibility. 
In quantum many-body systems, exponents of such divergence 
is known to follow the universality that has one extra dimension higher than the space dimension, 
and this additional dimensional degree of freedom represented by the imaginary time axis 
is responsible for quantum fluctuation. 
At low energies or low temperatures, the field theory gives a good description of the states near
the quantum critical point(QCP)\cite{Sachdev1999}. 
The knowledge about static criticality is thus established in both quantum and classical systems, 
providing reasonable interpretations 
to the experimental observations in laboratories\cite{Kono2015}.
\par
However, regarding the dynamics, 
how the physical properties react to the enhanced quantum fluctuation near the QCP remains unexplored. 
The difficulty stems primarily from a lack of theoretical tools for evaluating linear response functions 
in quantum many-body systems \cite{Kubo1957}. 
Although it is naively expected that the dynamical exponents will also follow the universality 
with one extra dimension, quantum relaxation processes remain hard to access even numerically. 
\par
In experiments, the dynamical response measurement in an applied field is a very useful technique. 
Observations at very low temperatures that appears to be influenced by quantum criticality 
have been reported from time to time, 
while unfortunately, they cannot be understood within the framework of available theories. 
One of the intriguing examples is the anomalous dielectric response in a series of triangular lattice 
Mott insulators, $\kappa$-(ET)$_2X$, $X$=Cu$_2$(CN)$_3$\cite{Abdel-Jawad2010} and Cu[N(CN)$_2$]Cl\cite{Lunkenheimer2012}. 
In these materials, the ET molecules are structurally dimerized and form 
a triangular lattice in the two-dimensional (2D) conducting layer as shown in Fig.~\ref{f1}(a). 
Each dimer accommodates a single charge in a Mott insulating phase at low temperature due to strong intra-dimer electronic
correlations\cite{Kanoda2006}. 
The former material possibly hosts a quantum spin liquid state in the same Mott insulating phase\cite{Shimizu2003}. 
Deep inside this phase, the temperature dependent dielectric function shows a peak at 
$T_m(\omega)$ which shifts significantly to lower temperature as the frequency $\omega$ is varied\cite{Abdel-Jawad2010}. 
Although such behavior is reminiscent of relaxer ferroelectrics found typically in PMN\cite{Vugmeister1997}, 
the frequency range where the peak shift is observed is much wider, varying over more than two orders of magnitudes. 
Physically, the peak temperature roughly corresponds to the energy scale dominating the system, 
and a single divergent peak structure generally suggests a ferroelectric phase transition at that temperature. 
The observation of frequency-dependent non-divergent peaks indicates a coexisting broad-range distribution 
of characteristic time and energy scales. 
In relaxer ferroelectrics, this phenomenon had been attributed to the polar-nano region 
induced by the artificial impurity doping\cite{Vugmeister1997,Mori2012}. 
However, the organic crystals are almost free of impurities. 
\par
The Mott dielectrics in organic crystals are attributed to the quantum electric dipole\cite{Chisa2010} 
-- the degree of freedom of charge to stay at either of the dimerized two molecular orbitals. 
A good description of this degree of freedom is provided by the transverse Ising (TRI) model\cite{Chisa2010,Naka2010}, 
a canonical model of quantum computation /annealing\cite{Kadowaki1998,Brooke1999} as well 
as of condensed matter theory. 
Each charge fluctuates back and forth within the dimer by quantum tunnelling (transfer integrals) 
as shown in Fig.~\ref{f1}(a), namely a transverse electric field is placed on the dipole 
and the Coulomb interactions between the charges (dipoles) are the Ising interactions. 
If they align in the same direction, they yield a quantum ferroelectricity (see Fig.~\ref{f1}(b)). 
The question is, what could be the reason for the coexisting massive range of energy scales in a uniform system at low
temperature?, and would it be clarified by the microscopic calculation on the TRI model 
without the aid of simplified phenomenology?\cite{Kishine2017}
\par
We construct a kinetic protocol based on Glauber dynamics using the quantum Monte Carlo (QMC) method, 
and obtain a dynamic susceptibility, $\chi(q=0, \omega)$, of the TRI model. 
We extract the relaxation timescale $\tau$ from the Monte Carlo dynamics and show that 
$\chi(q=0, \omega)$ turns out to be the Debye function about $\omega$ at fixed $k_BT$ 
whose half-width is given by $\tau^{-1}$. 
Since both $\tau$ and $\chi(q=0,\omega=0)$ diverge toward QCP in lowering the temperature, 
the peak narrowing occurs. 
This $\chi(q=0, \omega)$, when viewed as a function of temperature for fixed $\omega$ takes 
a maximum at $T_m(\omega)$ which significantly decreases with $\omega$ due to the peak-narrowing effect. 
Since $\chi(q=0, \omega)$ corresponds to the dielectric function of quantum electric dipoles, 
the aforementioned experimental observation can be understood as the signature of dynamical quantum criticality
in the vicinity of the charge ordering transition. 
\section{Model and Formulation}
\subsection{Transverse Ising model}
Let us introduce the TRI model in a two-dimensional anisotropic triangular lattice;
\begin{equation}
  \label{eq:1}
  H = \sum_{\langle i,j \rangle} -J_{ij}\sigma_i^z\sigma_j^z - \Gamma \sum_i \sigma_i^x. 
\end{equation}
The $z$-component of the Pauli operator, $\sigma_i^z = \pm 1$, accounts 
for the location of charges in the $i$-th lattice site representing a dimer, 
which we call either ``pseudo-spin" or ``quantum electric dipole". 
The transverse field, $\Gamma$, flips the pseudo spins up and down via 
$\sigma_i^x = \sigma_i^+ + \sigma_i^-$ where $\sigma_i^{\pm}$ is the raising and lowering operators. 
We consider the Ising interactions between quantum dipoles, $J_{ij}$, on neighboring dimers, $i$ and $j$. 
In the anisotropic triangular lattice, we take $J_{ij}=J$ and $J'$ for the bonds along 
the two directions and the rest, respectively, as shown in Fig.~\ref{f1}(b). 
We take ferromagnetic $J(>0)$ while vary $J'$ from antiferromagnetic to ferromagnetic values. 
This model is obtained by the strong coupling perturbation theory at the lowest order 
from the so-called extended Hubbard model\cite{Chisa2010}, a basic model of $\kappa$-(ET)${}_2X$,
which includes the on-site and inter-site Coulomb interaction between 
electrons and the transfer integrals. 
\par
Different configurations of electric dipoles on neighboring dimers 
have different Coulomb energies, which is the origin of $J_{ij}$ (Appendix \ref{app:1}). 
From the first principles calculation, the actual parameter values of the extended Hubbard model are precisely
evaluated\cite{Koretsune2014,KNakamura2009,Jeschke2012}, 
and we transform it to our $J_{ij}$ and $\Gamma$ (see Appendix \ref{app:1}). 
We could thus access the experimentally observed phenomena without bias or assumption 
by referring to our numerical results with these material parameters.
\par
The dynamical response to spatially uniform external field $h(t)$, 
represented by the perturbation $H'(t) = -\sigma_i^z h(t)$ added to Eq.(\ref{eq:1}), 
is calculated by the Kubo formula\cite{Kubo1957}. 
The susceptibility for wave number $q$ and frequency $\omega$ is given as 
\begin{equation}
  \label{eq:2}
  \chi(q,\omega) = \chi(q,0) + i\omega \int_0^\infty dt e^{i\omega t}
  \Psi(q,t)
\end{equation}
which is interpreted in the experiments as a dielectric function, 
$\epsilon(q,\omega) /\epsilon_0 = 1+\chi(q,\omega)$ ($\epsilon_0$ is
the permittivity of free space) in an applied electric field. 
Here, $\Psi(q,t)$ is the relaxation function given 
in an imaginary time($\tau$) and real time($t$) connected form as 
\begin{equation} 
  \label{eq:3}
  \Psi(q,t) = \int_0^\beta d\tau \langle \sigma_{-q}^z(i\hbar\tau)\sigma_q^z(t)\rangle, 
\end{equation}
where $\sigma_q^z(t) = e^{\frac{iHt}{\hbar}}(\sum_j\sigma_j^ze^{-iqr_j})e^{-\frac{iHt}{\hbar}}$
is the interaction picture of the Ising operator of wave number $q$. 
The imaginary time $\tau$ that appears as parameter $i\hbar\tau$ in Eq.(\ref{eq:3}) 
runs from zero to inverse temperature $\beta = (k_BT)^{-1}$. 
Since we consider the ferroelectric order of the quantum dipoles, 
we focus on the case of $q=0$ in the following. 
\par
Conventionally, Eq.(\ref{eq:3}) is calculated using the finite temperature Green's function. 
There, one performs the analytic continuation from $\tau$ to $t$, 
but it is reliable enough only when the analytic form of Green's function is available, which is not the case for 
strongly correlated quantum systems\cite{Sachdev1999}. 
Tracking real-time dynamics using numerical time evolution is limited to very small 
system sizes in the exact diagonalization, and to one dimensional 
system by the density matrix renormalization group\cite{White2004} and matrix 
product construction\cite{Vidal2004}, which allows for only short timescales. 
One of the authors developed the nearly exact dynamics of the thermal pure state 
for a long enough timescale\cite{Endo2018}, but is applied so far for $N\lesssim 30$. 
Recently, the dynamics of the imaginary time evolution is examined in the quantum Monte Carlo study\cite{Syljuasen2008,Grandi2011}, 
which illustrates that the nonadiabatic quantum dynamics at a leading order could be similar to 
the real-time ones\cite{Grandi2011}. 
The generalized dynamical scaling of the susceptibility-like quantity obtained averaged along the imaginary
time shows a good collapse\cite{Liu2013}.
\begin{figure*}[t]
  \centering
  \includegraphics[width=165mm]{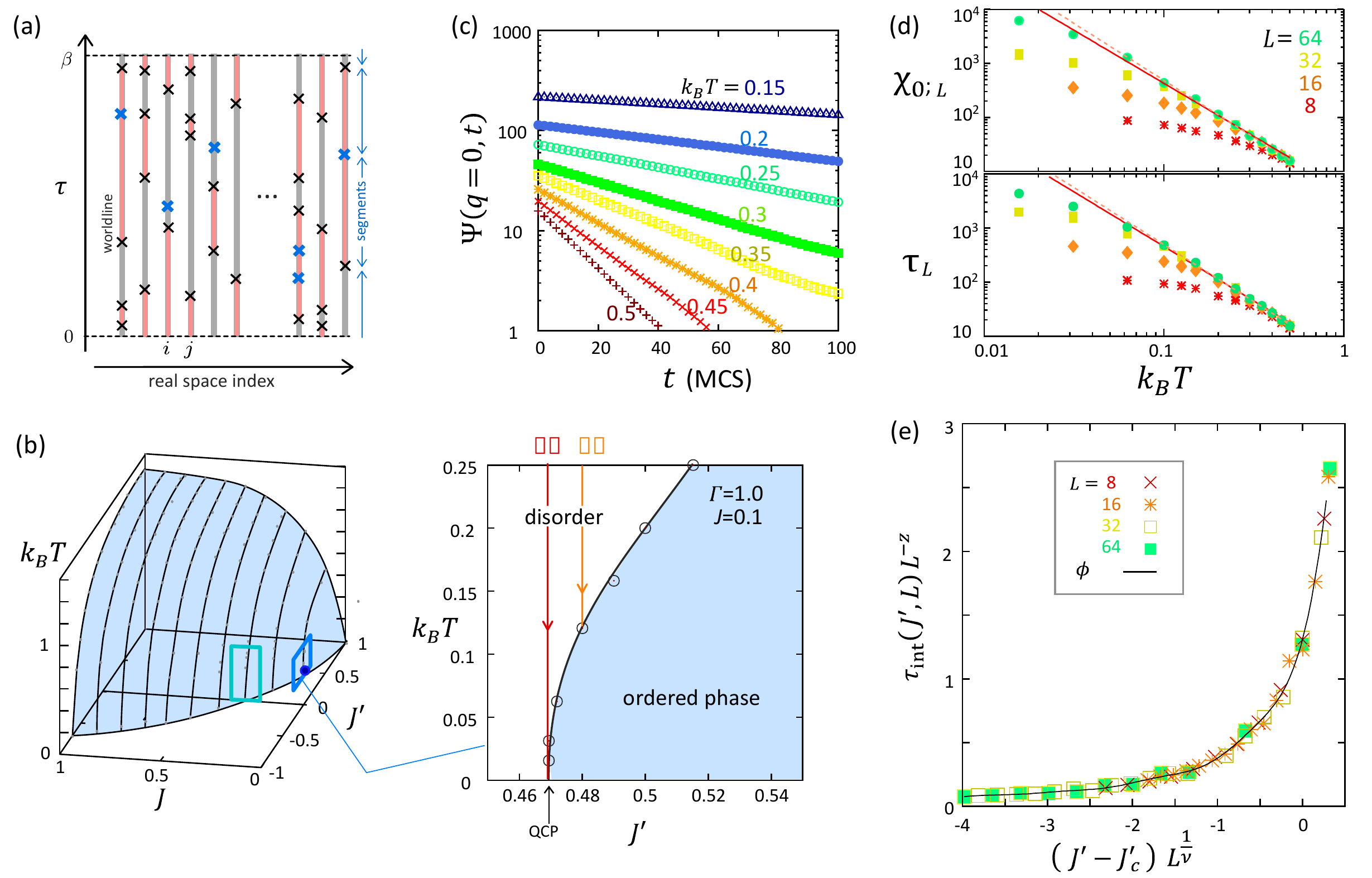}
  \caption{Results of QMC for the TRI model on an anisotropic triangular lattice given in unit of $\Gamma=1$. 
    (a) Schematic illustration of a set of world lines describing the partition function of the TRI model 
     given on the place of space($i$) and imaginary time($\tau$), which are periodic about $\tau=[0:\beta]$. 
     Kinks(cross symbols) including old(black) and new(blue) ones separate 
     the world lines into segments and the highlighted/plain segments 
     carry $\sigma^z=1/-1$. 
    (b) Phase diagram on the plane of $J$, $J'$, and $k_BT$. The shaded region corresponds to the
    ordered phase (ferroelectric order of dipoles). 
    The material parameters of $\kappa$-(ET)${}_2X$ (Appendix \ref{app:1}) fall near the blue point
    $(J\sim 0.1, J'\sim 0.5)$ in the diagram at $k_B T=0$. 
    The right panel shows the cross-section of the phase diagram at $J=0.1$ at low $k_BT$ in the vicinity of QCP. 
    The case of square lattice ($J'=0$) in green cross section is given in Appendix \ref{app:3}. 
    (c) Relaxation function $\Psi(q=0, t)$ obtained by the QMC calculation at $L=64$ and $J'=0.47, J=0.1$ for several
    choices of $k_BT$. 
    (d) $\tau_L$ and $\chi_{0;L}$ extracted from the relaxation function at several $L$, plotted as functions of
    $k_BT$. The envelope line (broken line) $\tau = c_1(k_BT)^{-z}, \chi_0=c_2(k_BT)^{-\frac{\gamma}{\nu}}$ 
    is their thermodynamic limit. The solid line represents the same function using the 3D critical exponents 
    and $c_1=4.34, c_2=4.61$, which is almost the
    same with the case of square lattice (Appendix \ref{app:3}, Fig.~\ref{figs2}). 
    (e) Dynamical finite size scaling analysis. Correlation time
    $\tau_{\rm int}$ is obtained for a series of $L=8,16,32$ and $64$
    down to $k_BT = 0.0078125J$ with $\Gamma/k_BT=2L$. The data
    collapse to a single scaling function $\phi$.
    }
  \label{f2}
\end{figure*}
\subsection{Kinetic protocol}
Traditional statistical mechanics has provided an idea to implement the dynamics in classical models; 
it is to consider an isolated system and observe the process of relaxation toward {\it local equillibrium} 
during {\it ``the time evolution"}. 
Glauber dynamics is one such realization using the Markov process\cite{Glauber1963}; 
when you apply the Markovian update of the state, a single target spin is locally relaxed quite 
immediately through the interaction with its surrounding spins that serve as a heat bath. 
Then, {\it ``the time evolution"} using the stochastic process, regardless of whether it is a heat bath method, 
Metropolis method or its analogs, was proved to reproduce well the critical behavior, 
where both static and dynamical exponents are successfully extracted. 
This was possible because the energetics is determined strictly locally 
in the classical system with short-range interaction, which does not apply to quantum systems in general. 
\par
However, the TRI model exceptionally realizes a quantum local equilibration, 
to which we can apply the idea of Glauber dynamics. 
Let us first overview the quantum Monte Carlo description of the TRI model. 
The partition function of the TRI model appears to be the ensemble of world lines running along the imaginary time direction $\tau=[0:\beta]$ 
with a periodic boundary, as shown in Fig.~\ref{f2}(a). 
Since we take the quantization axis parallel to $\sigma_i^z$, 
each point along the $i$-th world line takes either $\sigma_i^z=\pm 1$, and interacts by $J_{ij}$ with the pseudo-spins on 
the neighboring $j$-th world line. 
The quantum fluctuation represented by the transverse field works independently for each site-$i$, 
and when the pseudo-spin flips at some imaginary time $\tau$, it is represented by the kink on the world line. 
The kinks are inserted stochastically following the Poisson distribution 
and separate the world lines into segments. 
The weight each segment carries is the integrated classical Boltzmann weight about the Ising interaction with the neighboring pseudo-spins at the same $\tau$. 
The Markov process is summarized into the following steps; 
\begin{itemize}
\vspace{-2mm}
\item[1] Choose site $i$ to update,
\vspace{-2mm}
\item[2] Stochastically generate a series of new kink-candidates along the $i$-th world line via 
Poisson process with $\Gamma$,
\vspace{-2mm}
\item[3] Separate the world line into segments by old kinks and kinks-candidates, 
\vspace{-2mm}
\item[4] Update $\sigma^z_i$ on each segment $\tau \in [{\tau_s}:{\tau_f}]$ following the thermal-bath method using the weight, 
${\rm exp} (\int_{\tau_s}^{\tau_f} \sum_{j} J_{ij}\sigma_i^z(\tau)\sigma_j^z(\tau) d\tau)$. 
\vspace{-2mm}
\item[5] We repeat these steps for $i\in [1:N]$. 
\vspace{-2mm}
\end{itemize}
The segments are locally updated independently of the rest of the system other than its neighboring segments, 
which produces the situation of the classical Glauber dynamics. 
Namely, the above-mentioned Markov process safely relaxes the TRI model toward thermal equilibrium 
by making use only of the local updates in a unit of segments. 
Importantly, this process was empirically proved to successfully reproduce the dynamical scaling relation 
of the TRI model on the square lattice\cite{TotaNakamura2003}. 
By taking $\Gamma\rightarrow 0$, we find the smooth connection to the Glauber dynamics of the classical Ising model. 
\par 
We study the dynamical properties using this Markov process which we call a kinetic TRI protocol. 
The evaluation of Eq.(\ref{eq:3}) is straightforward. 
We approximate the two time evolutions to be independent and 
denote the two variables explicitly as, $\sigma^z(\tau,t)$, 
where the real-time $t$ is the Monte Carlo step. 
We measure $\langle \sigma^z_j(\tau,s) \sigma^z_i(0,s+t)\rangle_{\rm eq}$ 
between $\sigma_i^z$ of $t=s$ at imaginary time $\tau$, 
and that of $t=s+t$ and imaginary time $0$, where the integration of $\tau=0\sim\beta$ 
is made independent of $t$. 
We take an average over $M$ time-steps in the equilibrium as; 
\begin{eqnarray}
\Psi(r_i-r_j,t)&=& \int_0^\beta d\tau \big\langle\; 
 \sigma^z_{j}(\tau,0) \sigma^z_{i}(0,t)\;\big\rangle_{\rm eq} 
\nonumber\\
&=&\frac{1}{M} \sum_{s=0}^M \big\langle\bigg(\int_0^\beta d\tau \sigma^z_{j}(\tau,s) \bigg) \sigma^z_{i}(0,s+t) \big\rangle
\\
\Psi(q,t)&=& \frac{1}{N} \sum_{i=1}^N \sum_{j=1}^N e^{-i q (r_i-r_j)} \Psi(r_i-r_j,t).
\label{eq:psi}
\end{eqnarray}
Our QMC calculation is performed for a $N=L\times L$ site cluster 
with $L=8, 16, 32, 64, 128$, while taking $L \times k_BT = 8, 4, 1, 0.5$. 
This is because near the QCP, the minimum temperature that captures the relatively size-free $(L > \xi)$
results is limited at each $L$, and the correlation length $\xi$ diverges in powers. 
Similarly, the time correlation represented by the relaxation time $\tau_L$ extends to more than 10$^6$ steps near QCP,  
so that we averaged Eq.(\ref{eq:psi}) over 16 runs, taking $M=10^7$ time steps for each. 
\par
Finally, we notice that some other protocols are applied to quantum annealing\cite{Martonak2002,Heim2015}, 
while they do not fulfill the condition for Glauber dynamics; Ref.[\onlinecite{Heim2015}] includes the loop update 
and Ref.[\onlinecite{Martonak2002}] performs simultaneous flipping of a variable along the whole imaginary time. 
Particularly in the latter the relaxation process may change and shall be discriminated from Ref.[\onlinecite{TotaNakamura2003}]. 
We briefly note that there are some other trials like a phenomenological extension of
the Glauber dynamics to quantum systems\cite{Yin2016}, or variational Monte Carlo approaches 
regarding time evolutions\cite{Blas2016}, and semiclassical approximation using the discrete Monte Carlo
sampling in phase space\cite{Schachenmayer2015}. 
\par
\begin{figure*}[th]
  \centering
  \includegraphics[width=180mm]{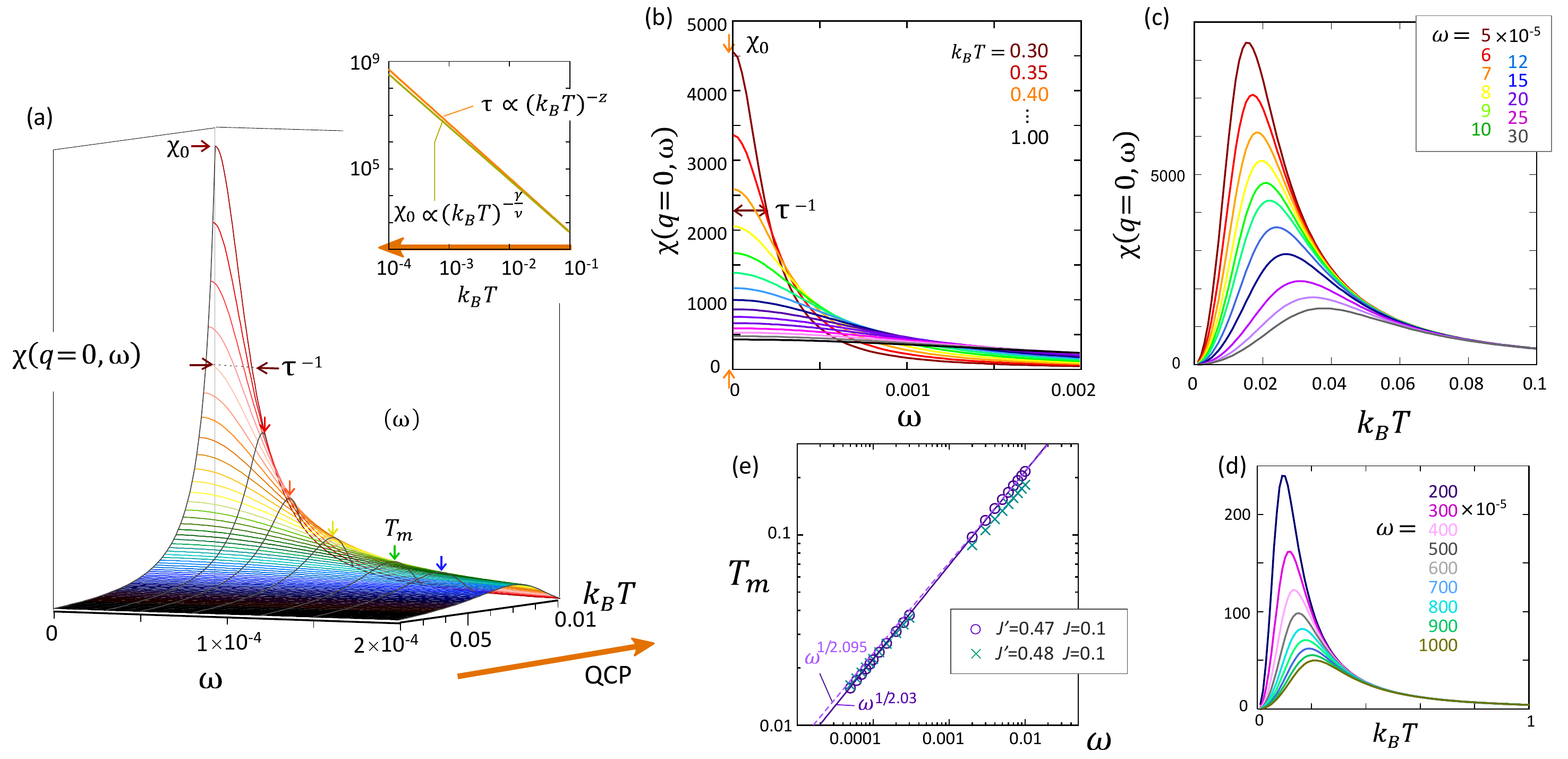}
  \caption{Dynamical susceptibility $\chi(q=0,\omega)$ obtained by the kinetic TRI protocol 
   as functions of $\omega$ and $k_BT$. 
   (a) Three-dimensional plot showing a set of Debye functions about $\omega$ 
   for different choices of $k_BT$. We choose $(J',J)=(0.47,0.1)$ that exhibits QCP in 
   Fig.~\ref{f2}(b).  
   Inset shows the temperature dependence of $\tau$ and $\chi_0$ used for this plot 
   and are extracted from the relaxation function. 
   Small arrows indicate the peak positions, $T_m(\omega)$. 
   (b) $\omega$-dependence of $\chi(q=0,\omega)$ for several choices of $k_BT$. 
   The half-width of the $q=0$ peak gives $\tau^{-1}$ and its peak height gives $\chi_0$. 
   (c,d) Temperature dependence of $\chi(q=0,\omega)$ for several choices of $\omega$, 
         ranging over $5-30 \times 10^{-5}$ and $2-10 \times 10^{-3}$. 
   (e) $\omega$-dependence of $T_m$ for the QCP data(in panel (c,d)) 
    and for slightly off QCP, $(J',J) = (0.48,0.1)$. Solid and broken lines are 
    $\propto \omega^{1/2.03}$ fitted by the QCP data and $\omega^{1/z}, z=2.095$, respectively. 
 }
  \label{f3}
\end{figure*}
\begin{figure}[th]
  \centering
  \includegraphics[width=70mm]{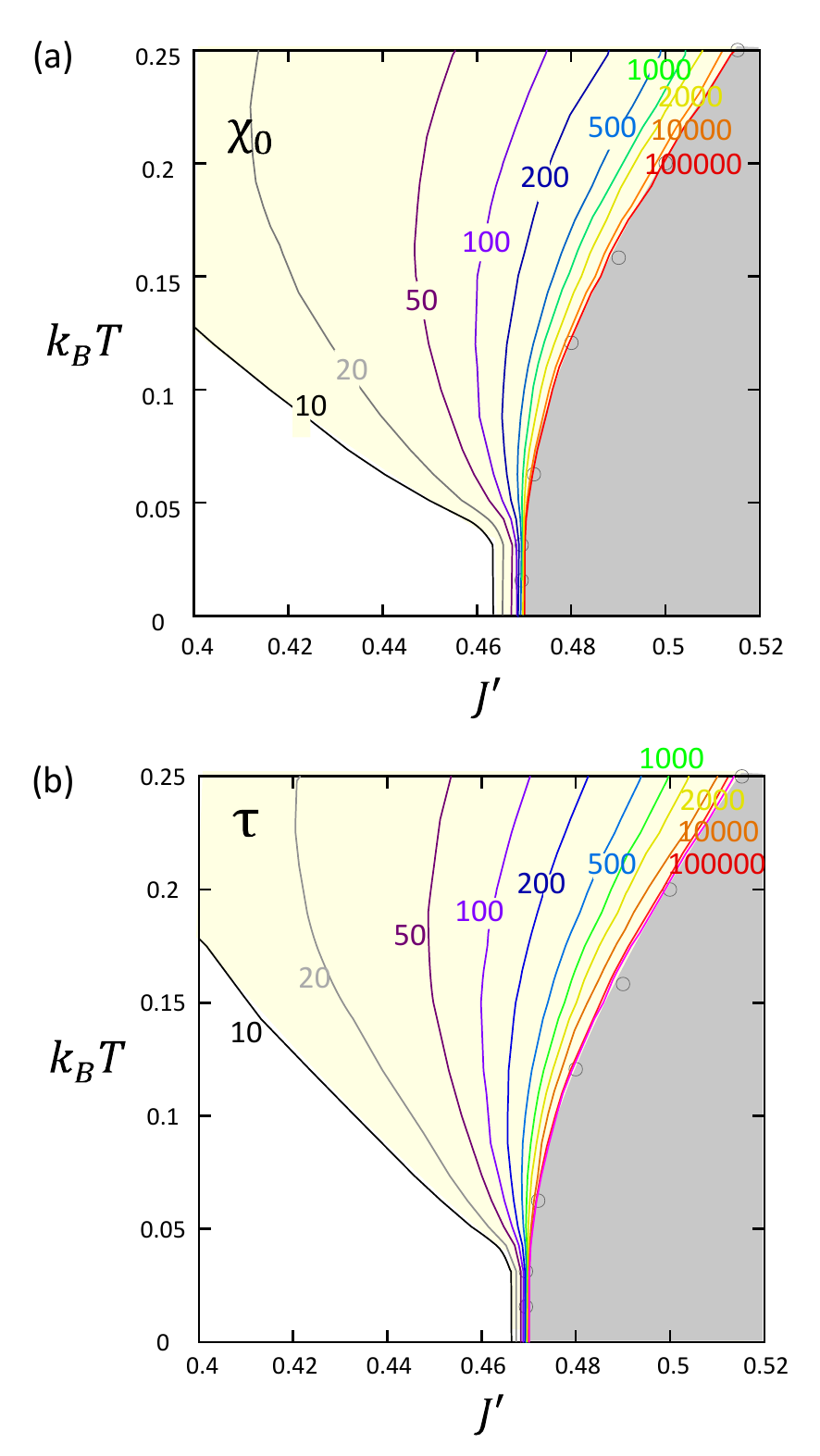}
  \caption{Density map of (a) static susceptibility $\chi_0$ and (b) the relaxation time $\tau$ extracted 
    from the relaxation function at several $L$, plotted as functions
    of $k_BT$ for the TRI model on a triangular lattice taking $\Gamma=1$ as a unit.}
  \label{f4}
\end{figure}
%
%
\section{Results}
\subsection{Phase diagram}
We first overview the low-temperature properties of the TRI model on an anisotropic triangular lattice. 
Overall, at large enough $J_{ij}/\Gamma$ the system is in an ordered phase, 
while the increase of $\Gamma$ makes the system disordered, 
and the phase transition between the two is typical second-order. 
We show the $k_BT - J - J'$ phase diagram in Fig.~\ref{f2}(b) in unit of $\Gamma=1$ 
obtained by the present QMC calculation. 
We made a Binder plot of the pseudo-spin expectation value $\langle \sum_j \sigma_j^z \rangle$ to evaluate the 
phase boundary and compared it with the anomaly of the specific heat, which turned out to be consistent. 
\par
The ordered phase extends from the large $J,J' > 0$ region toward slightly 
antiferromagnetic $J'$. The case of the square lattice ($J'=0$) is 
well studied\cite{TotaNakamura2003,Ikegami1998} and the phase boundary at $k_BT= 0$(QCP) is evaluated as,
$J_c/\Gamma = 0.3284(9)$\cite{TotaNakamura2003}. 
From a series of first-principles calculations, a family of $\kappa$-(ET)${}_2X$ is located at around
$J\sim 0.1, J' \sim 0.5$ (Appendix ~\ref{app:1}) \cite{Koretsune2014}, which is marked
in Fig.~\ref{f2}(b). One finds that it is near the QCP. 
%
%
\subsection{Relaxation function}
In the disordered phase relatively near the phase boundary, 
the relaxation function $\Psi(q=0, t)$ shows a clear exponential decay as a function of QMC time step 
typically as in Fig.~\ref{f2}(c), which can be described as 
\begin{equation}
  \Psi(q=0, t) = \chi_{0;L} \exp(-t/\tau_L),
  \label{eq:3a}
\end{equation}
using the static uniform susceptibility, $\chi_{0;L}$, and the relaxation time $\tau_L$ 
at fixed $k_BT$, $J, J'$ and $L$. 
The extracted values of $\chi_{0;L}$ and $\tau_L$ are plotted in Fig.~\ref{f2}(d) 
for $L=8, 16, 32$ and $64$ as functions of $k_BT$ at $J'=J'_c$. 
Data points belonging to different $L$ follow different curvatures, 
which converge to an envelope function given in a solid line: they are the values at the thermodynamic limit, 
which we denote $\tau$ and $\chi_0$. 
When the correlation length $\xi$ exceeds $L$ at low $k_BT$, the data points fall off from the envelope function. 

\subsection{Finite size scaling analysis}
We now test the similarities between the present kinetic TRI protocol and the original TRI model 
by the generalized dynamical finite-size scaling analysis; 
the scale invariance is expected in the dynamical critical phenomena, which results in the finite size scaling 
form of the relaxation timescale near QCP given as, 
\begin{equation}
  \label{eq:6}
  \tau_{\rm int}(J', L) = L^z \phi((J'-J_c')L^{\frac{1}{\nu}}), 
\end{equation}
where $z$ is the dynamical critical exponent and 
$\nu$ is the critical exponent characterizing $\xi\propto |J'-J'_c|^{-\nu}$. 
We evaluate $\tau_{\rm int}$ at low temperatures available in a series of $k_BT=\Gamma/2L$ 
down to $k_BT = 0.0078125\Gamma$ with $\Gamma=1$ 
by varying $J'$ in the phase diagram of Fig.~\ref{f1}(b). 
We use the following integral,
\begin{equation}
  \label{eq:7}
  \tau_{\rm int}=\int_0^\infty \Psi(q=0,t)\Big/\Psi(q=0,0) dt
\end{equation}
which gives the value independent of the detailed functional form of
$\Psi(q=0, t)$. Figure~\ref{f2}(e) shows the finite size scaling plot using
$L=8,16,32$ and $64$. One finds an almost perfect collapse of the data
points into a single functional form. The exponent obtained by this
plot is $J_c/\Gamma=0.4700, (z, 1/\nu)= (2.095,1.56(3))$, which is 
fully consistent with our Binder analysis of TRI and the fitting of
exponents on the kinetic TRI. 
We thus think it to be properly interpreted as a 3D universality class. 

\subsection{Susceptibility and critical exponents}
We have shown that the relaxation function decreases exponentially with $t$ as Eq.(\ref{eq:3a}), 
and one can extract from a series of $\chi_{0;L}$ and $\tau_L$, their $L\rightarrow\infty$ limit, $\tau$ and $\chi_0$. 
Since the system is near QCP, $\tau$ and $\chi_0$ diverge in powers 
toward the ordered phase as (see the inset of Fig.~\ref{f3}(a)), 
\begin{equation}
  \tau(k_BT) = c_1(k_BT)^{-z},\;\; \chi_0(k_BT)=c_2(k_BT)^{-\frac{\gamma}{\nu}}
  \label{eq:4}
\end{equation}
where $\gamma$ is the magnetic critical exponent and $c_i$ are the constant coefficients. 
This could be understood as follows; Consider a quantum 2D system of size 
$L\times L$ with an additional axis in the imaginary time direction, $[0:\beta]$, 
that characterizes the quantum fluctuation. 
As the system approaches QCP, the correlation length $\xi$ diverges. 
Suppose that $L$ is large enough to assume $L > \xi$, 
and then $\beta$ becomes the upper bound of the effective system length. 
For moderately low temperatures, $\xi$ cannot develop larger than $\beta$. 
From the scaling theory, we immediately find $\tau \propto \xi^z = (k_BT)^{-z}$. 
The form Eq.(\ref{eq:4}) is applied to laboratory systems as well as to theoretical models. 
\par
The criticality at QCP and off QCP in the ordered region follows that of the 3D and 2D (kinetic) Ising universality classes\cite{Cardy1996,Sachdev1999} 
(the two lines in the right panel of Fig.~\ref{f2}(a)), and their exponents are evaluated as
$(z,\gamma,\nu) =$ 
(2.02\cite{Collura2010}-2.03\cite{Wansleben1987},1.237,0.629\cite{Gliozzi2014}-0.630\cite{Hasenbusch2010}) and
(2.165\cite{Ito1993}-2.18\cite{Stauffer1992,Dammann1993},1.75,1), respectively. 
We analyzed the QMC data precisely and found good agreement with these exponents(Appendix \ref{app:3}, Fig.~\ref{figs2}). 
The envelope of Fig.~\ref{f2}(d) follows these exponents. 
It is notable that $c_1$ and $c_2$ does not
seem to depend on the location of QCP in the phase diagram. 
\par
The dynamical susceptibility in Eq.(\ref{eq:2}) is a Fourier transform of 
Eq.(\ref{eq:3a}), which is given analytically in the Lorentzian form as,
\begin{equation}
  \chi(q=0,\omega)=\chi_0\frac{\tau^{-2}}{\omega^2+\tau^{-2}}.
  \label{eq:5}
\end{equation}
It corresponds to the Debye function in dielectrics. The cross-sections
of Fig.~\ref{f3}(a) at fixed values of $\omega$ and $k_BT$ are
shown in Fig.~\ref{f3}(b) and Fig.~\ref{f2}(c), respectively. 
The frequency dependence of $T_m(\omega)$ near QCP is scalable, 
namely, if we take the temperature range one order of magnitude higher than that 
of the main panel of Fig.~\ref{f2}(e), the almost same functional form is 
observed by shifting the frequency to the higher energy (Fig.~\ref{f3}(d)). 
\par
Let us apply the scaling analysis to the dynamical susceptibility. 
Reminding the form of $\chi_0$ in Eq.(\ref{eq:4}) at $|J'-J'_c|\rightarrow 0$, 
one can express Eq.(\ref{eq:5}) as $\chi(T,\omega)= T^{-\gamma/\nu} \psi (\omega\tau)$.  
In finite size systems, in approaching QCP 
the correlation length cannot exceed $\xi \sim \beta$ and accordingly, 
$\tau \propto \xi^z \sim T^{-z}$, 
which means that $\chi(T,\omega)= T^{-\gamma/\nu} \psi (\omega T^{-z})$. 
The peak position of this function fulfills 
\begin{equation}
T_m \propto \omega^{1/z}. 
\label{eq:tmomega}
\end{equation}
The data points shown in Fig.~\ref{f3}(e) obtained from Figs.~\ref{f3}(c) and \ref{f3}(d) 
indeed follow this power-law dominated by the dynamical critical exponent. 
As we discuss shortly, this behavior is in good agreement with the dielectric experiments 
on $\kappa$-ET$_2$Cu$_2$(CN)$_3$. 
\par
By precisely evaluating Eq.(\ref{eq:3}) by the QMC calculation 
and from the size scaling, we obtain a set of $(\chi_0,\tau)$ in Eq.(\ref{eq:5}) 
over the whole region of the phase diagram. 
Their contour maps are given in Fig.~\ref{f4}. 
One can regard the region $\tau < 10$ of being no longer
critical, namely either quantum mechanically or classically
disordered, which is marked as a region outside the red line in
Fig.~\ref{f1}(b) (For the corresponding actual value of the square lattice,
see Appendix B, Fig.~\ref{figs2}). The naive and schematic description of
the crossover lines of the QCP region are generally given as in the
yellow broken line of Fig.~\ref{f1}(b), which however turned out to be only
qualitative and valid at extremely low temperatures.

\section{Summary and Discussion}
We have clarified how quantum criticality appears in the dynamics of the quantum many-body system. 
As an ideal and realistic platform, we chose the transverse Ising model on an anisotropic triangular lattice, 
whose Ising degrees of freedom represent the quantum electric dipole degrees of freedom 
in the dimer Mott insulating phase of the organic crystal, $\kappa$-ET$_2X$. 
The same set of calculations is also performed for the case of the regular square lattice 
to confirm that the results are not dependent on the model parameters. 
The model is known to exhibit a quantum criticality and can be almost exactly solved 
numerically by the quantum Monte Carlo method. 
We developed a kinetic TRI protocol to study the quantum dynamics of the transverse Ising model, 
which is built on the local quantum Monte Carlo update of segments of worldlines running in the 
imaginary time directions. 
This Markov update enables a rapid local equilibration of each segment, 
that can be mapped to the case of classical Monte Carlo updates of higher dimensions. 
Since the latter is known to capture the intrinsic real-time Glauber-type dynamics, 
our Monte Carlo time can mimic real-time relaxation in the same context, 
allowing us to study the semi-classical dynamics representing the quantum dynamics of the original model. 
\par
In this protocol, we obtained the dynamical susceptibility by analyzing the Monte Carlo time dependence of the correlation functions, 
and showed that they have Debye functional form 
with its peak heights and inverse of width diverging algebraically in approaching QCP. 
This led to a significant peak-narrowing and the obtained temperature dependence of the 
dynamical susceptibility is found to show a frequency-dependent peak shift, 
reminiscent of the relaxor-ferroelectric-like behavior observed in many experimental studies 
of organic dimer Mott materials\cite{majed2013,jens2020}. 
\par
We briefly refer to some theoretical studies discussing this relaxor-ferroelectric-like behavior of $\kappa$-ET$_2X$. 
The extended Hubbard model in one dimension is studied at the mean-field level using the phase Hamiltonian\cite{Kishine2017}, 
which they aim to represent phenomenologically the cross-section line of the two-dimensional systems. 
They discussed the kinks (the domains in 2D) as the origin of frequency-dependent peaks, 
and by evaluating the dynamical correlation function of kinks, 
showed that their relaxation timescale shall vary with frequency by orders of magnitudes. 
This may give one simplified interpretation of part of the phenomena. 
However, they do not explain a temperature-dependent characteristic dynamical susceptibility 
and phenomena seem to have no relevance to the criticality we observed. 
\par
The authors in Ref.[\onlinecite{Kennett2022}] have studied the analogue the effective model 
in Ref.[\onlinecite{Chisa2010}] for $\kappa$-ET$_2X$. 
Then, they discarded the quantum fluctuation term and performed the classical Monte Carlo study, 
where they took account of the electron spin as a classical SO(3) vector 
which coupled with the electric dipole described as as Ising pseudo-spins, showing that the two will generate 
a dynamical (classical) disorder to each other. 
The dipole susceptibility shows broad peaks in lowering the temperatures, 
which they attributed to the glassiness; it may be relevant to the glassy behavior of 
$\kappa$-ET$_2$Cu$_2$(CN)$_3$ at $T<6$K\cite{Abdel-Jawad2010}.
Indeed, the coupling of two different degrees of freedom can be a driving force of glassiness. 
Recently, one of the authors and collaborators showed that in a three-dimensional frustrated pyrochlore lattice, the model including the spin and lattice-displacement coupling can exhibit 
a thermodynamic glass transition at finite temperature even without quenched disorder\cite{Mitsumoto2020}, 
which explained the long-standing puzzle on the origin of the disorder-free spin glass 
in Yb$_2$ Mo$_2$O$_7$\cite{Gingras1997}. 
Since the classical model in Ref.[\onlinecite{Kennett2022}] is two-dimensional, 
the fluctuation disturbs the true glass transition and the system remains glassy. 
If one deals with it quantum mechanically, there shall be room for the true glass transition\cite{Chisa2022}. 
\par
We now compare the overall behaviour of $\chi(q=0,\omega)$ with the
experimentally observed\cite{Abdel-Jawad2010,Lunkenheimer2012} 
dielectric constant $\epsilon'(\omega)$ of $\kappa$-(ET)${}_2X$. 
The material at ambient temperature is a good conductor. 
At temperatures below 100K, the charges start to lose their conductance and localize on
each dimer, and a quantum electric dipole is spontaneously formed 
due to strong electronic interactions\cite{Chisa2010}. 
This electric dipole emerges due to the special modulation of 
wave function (charge distribution), 
which should be discriminated from the conventional and semiclassical lattice-displacement types of
dielectrics\cite{Khomskii2009}. 
As the frequency is varied from 1kHz to 100kHz the peak position of the dielecric constant of 
$\kappa$-(ET)${}_2$Cu${}_2$(CN)${}_3$ shifts from about 20K to 50K. 
By extracting $\epsilon'(\omega)$ within this temperature window 
and fitting them by Eq.(\ref{eq:5}), we find a series of Debye curves belonging to different $T$, 
that crosses in a manner comparable to Fig.~\ref{f2}(e) (Appendix \ref{app:5}). 
In the case of $\kappa$-(ET)${}_2$Cu[N(CN)${}_2$]Cl, only slight variation of 
$T_m(\omega) \sim 25-30$K is found, with no such crossings, and is considered to locate off the QCP.
\par
One remaining issue is that we cannot directly determine the laboratory timescale 
that corresponds to the Monte Carlo timestep. 
Still, we may safely assume that for each temperature, 
$t=a(T)t_{\rm lab}$, holds, where $a(T)$ could become smaller with 
lowering the temperature by a few factors. If we plot the extracted of
$\kappa$-(ET)${}_2$Cu${}_2$(CN)${}_3$ against $(T-T_c)$, taking $T_c=6$K
where the Curie tail of $\epsilon'(\omega)$ diverges\cite{Abdel-Jawad2010}, 
we obtain $\tau_{\rm lab} \propto (T-T_c)^{z_{\rm lab}\nu}$ with 
$z_{\rm lab}\nu \sim 2-3$ (Appendix Fig.~\ref{figs4}) not too 
different from that of 2D Ising ones expected for the case with 
finite.
\par
Although there had been a dispute on whether such seemingly subtle
dipole really exists\cite{Pinteric2014,Sedlmeier2012}, further examination on
$\kappa$-(ET)${}_2$Cu[N(CN)${}_2$]Cl after Ref.[\onlinecite{Lunkenheimer2012}] for many samples
supported the picture of the order-disorder type of ferroelectrics\cite{Lang2014}. 
The dipoles have further proven to be present 
in $\beta$-(ET)${}_2$ICl${}_2$, a similar 2D material showing the same critical dynamics, 
via observation of pyrocurrent\cite{Iguchi2013}, collective
mode \cite{Itoh2013}, and the polarization curve \cite{Hattori2017}. 
The noize measurements on $\beta$-(ET)${}_2$ICl${}_2$ suggests 
an emergent nanoscale polarized cluster\cite{jens2020} which is apparently not due to impurities. 
The phenomena is not restricted to ET systems is observed in another dimer Mott insulator, 
$\beta'$-type Pd(dmit)$_2$\cite{majed2013}. 
Similar dynamics is quite relevant near the phase transition in a series of
quasi-one-dimensional organic materials TMTSF${}_2$X \cite{Nad2000,Monceau2001} 
based on dimerized molecules, although its criticality was not really 
discussed before. 
\par
The quantum nature of dielectrics has become a topic in a series of
materials; A geometrical frustration-induced quantum paraelectric
nature is found in the conventional displacement-type of dipoles in a
hexaferrite BaFe${}_{12}$O${}_{19}$\cite{Shen2016}. Critical behavior of
the static dielectric function has been discussed in another
displacement-type of quantum paraelectric, SrTiO${}_3$, on the basis
of a phenomenological $\phi^4$theory which explains well the
experimental observation in such a three-dimensional system with
moderate quantum fluctuation\cite{Rowley2014}. Then finally, the present study
reached the dynamics of dipoles in the presence of strong quantum
fluctuation characteristic of two dimensions. The TRI model adopted
here may serve as an intersection of material science in laboratories
and the modern theories of computational science.

\begin{acknowledgments}
This work was supported by a Grant-in-Aid for Transformative Research Areas 
``The Natural Laws of Extreme Universe, A New Paradigm for Spacetime and Matter
from Quantum Information (Grant No. 21H05191) and 
other JSPS KAKENHI (No. 21K03440, 18H01173) of Japan.
We thank the experimentalists, 
  Takahiko Sasaki, Ichiro Terasaki, Jens Mueller, Peter Lunkenheimer,
  Michel Lang and Martin Dressel for fruitful communications. 
 We also thank Sei Suzuki for the discussions. 
\end{acknowledgments}

\appendix
\section{Microscopic derivation of the model parameters} 
\label{app:1}
\begin{figure}[tbp]
\begin{center}
\includegraphics[width=6.5cm]{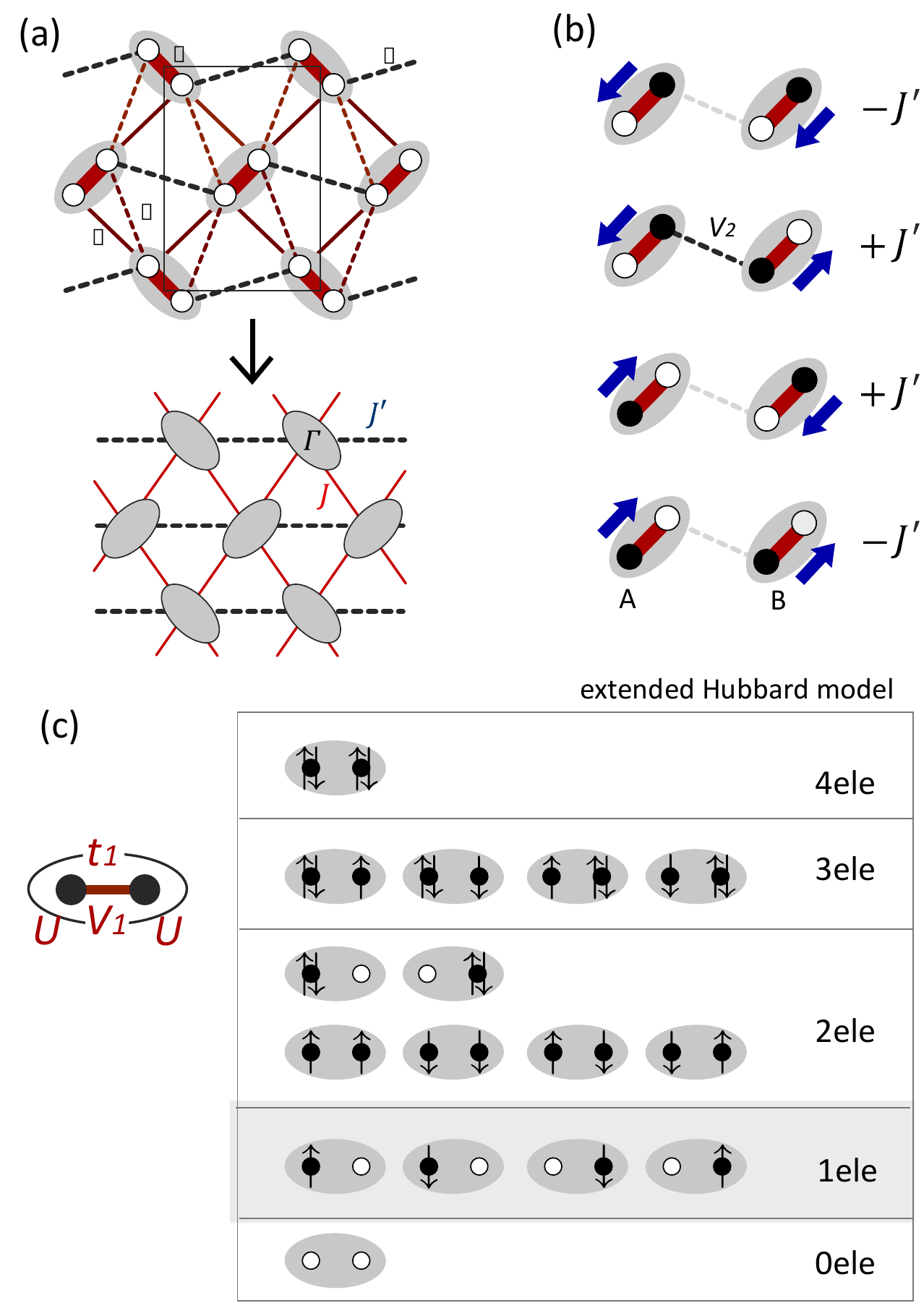}
\end{center}
\caption{Schematic description of the models of $\kappa$-ET$_2X$. 
(a) the mapping of the extended Hubbard model 
based on molecules(circles) to the transverse Ising model based on dimers(ovals). 
The indices on four independent bonds are those of $t_i, V_i$, $i=1\sim 4$. 
(b) The 16 basis of the extended Hubbard model on a single dimer, where $\uparrow$ and $\downarrow$ indicate 
the electrons of up and down spins, respectively. 
The four different configurations with one electron per dimer form the low energy local Hilbert space at large $V_1,U$. 
When the spin degrees of freedom are neglected at the leading order of perturbation, they are reduced to two. 
(c) Configuration of electrons on adjacent two dimers, where the arrows indicate the corresponding 
pseudo spin configuration. The second panel has energy $V_2$ and others zero, which yields the 
Ising interactions between pseudo spins, see the text. 
}
\label{figs1}
\end{figure}

We evaluate the model parameters of $\kappa$-ET$_2X$ based on the 
first principles calculation reported by one of the authors\cite{Koretsune2014}. 
Figure \ref{figs1} shows the schematic description of the two dimensional 
conducting layer of $\kappa$-ET$_2X$, 
where the circle represents an ET molecular orbital (we call here ``site") and the oval a dimer. 
There are four sites and two dimers in the unit cell. 
This family of material has an old history\cite{Kanoda2006}, 
and is well described by the extended Hubbard model in a unit of molecular orbitals as\cite{chisa-review}, 
\begin{equation}
{\mathcal H}=\sum_{\langle i,j\rangle}\sum_{\sigma=\uparrow,\downarrow} 
-t_{ij} \big(c_{i\sigma}^\dagger c_{j\sigma} + {\rm H.c.}\big)
+ \sum_{i=1}^N U n_{i\uparrow}n_{i\downarrow} 
+ \sum_{\langle i,j\rangle} V_{ij} n_in_j
\end{equation}
where $c_{i\sigma}^\dagger/c_{i\sigma}$ is the creation/annihilation operator 
of electrons on-site $i$ and spin $\sigma$, 
and $n_{i\sigma}=c_{i\sigma}^\dagger c_{i\sigma}$, $n_i=n_{i\uparrow}+n_{i\downarrow}$ 
are their number operators. 
The transfer integrals $t_{ij}$ are evaluated from the latest first principles calculation as 
(Table I and II of Ref.[\onlinecite{Koretsune2014}]), 
$(t_2,t_3,t_4)=(0.46,0.43,-0.08)$ and $(0.34,0.51,-0.21)$ 
for $X=$Cu$_2$(CN)$_3$ and Cu [N(CN)$_2$]Cl, respectively, 
in unit of $t_1$, showing that the geometry of $t$'s depends on materials. 
The intra-dimer transfer integral is not much different between materials; 
$t_1=199$ meV and 207 meV for $X=$Cu$_2$(CN)$_3$ and Cu [N(CN)$_2$]Cl, 
respectively. They take $195-209$meV for all other $\kappa$-ET$_2X$ studied in Ref.[\onlinecite{Koretsune2014}].
The on-site Coulomb $U$ and the inter-site Coulomb interactions $V_{ij}$ 
are also evaluated based on the molecular distances (X-ray structure)\cite{Koretsune2014} 
referring to the {\it abinitio} down-folding\cite{KNakamura2009}, which are 
$U=8$, $(V_1,V_2,V_3,V_4)=(4.0,2.0,2.4,2.0)$ in unit of $t_1=200$ meV (or $V_2\sim 0.4$eV), 
also almost independent of $X$. 
While the amplitudes of these interactions are overestimated, 
the ratio between these interactions shall be safely adopted. 

\begin{figure*}[t]
\includegraphics[width=12cm]{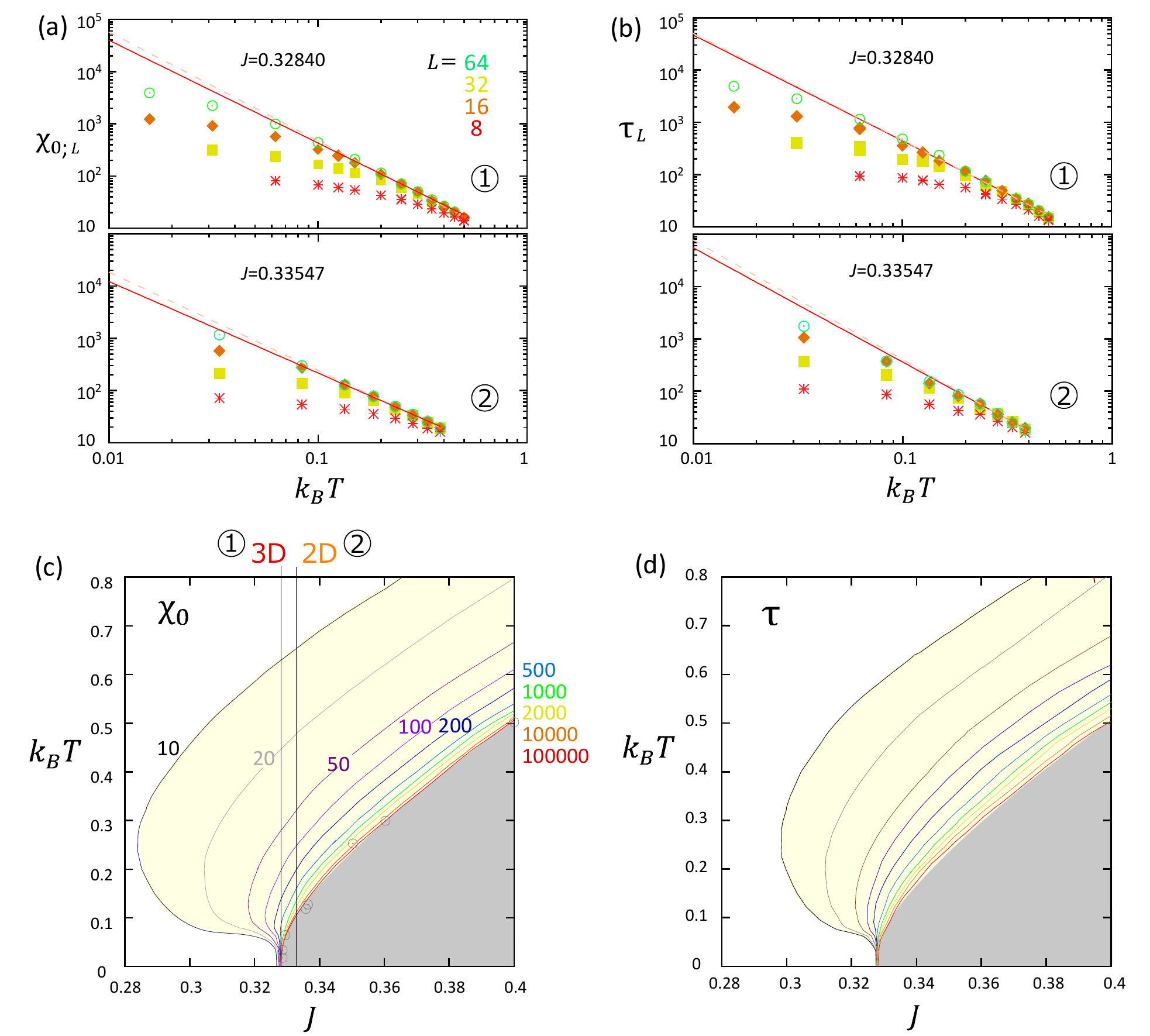}
\caption{Results of QMC calculations for the square lattice, $J'=0$. 
(a) Static susceptibility $\chi_{0;L}$, and (b) relaxation time, $\tau_L$, extracted from the relaxation function 
$\Psi(q=0,t)$ at $L=8,16,32,64$, plotted as functions of $k_B T$ for $J=0.32840$ and 0.33547 
at QCP and slightly off QCP. Error bars are smaller than the symbols. 
The solid ad broken lines follow $\tau=c_1(k_BT)^{-z}$ and $\chi_0=c_2(k_BT)^{-\gamma/\nu}$. 
The red solid lines follow the critical exponent 
$(z,\nu,\gamma)=(2.02,0.629,1.2379)$ and (2.183,1,1.75) for $J'=0.32840$(top) and 0.33547(bottom), 
of the 3D and 2D Ising universality class, respectively, where the former(top panel) yields, 
$c_1=4.225,c_2= 4.623$ which is similar to the case of the anisotropic triangular lattice in the same phase diagram in the main text. (Fig.2b). 
Broken lines are the fitted envelope functions describing the thermodynamic limit 
with an exponent of $(z,\gamma/\nu)=(2.04,1.87)$ and (2.11,1.74) in the top and bottom panels, respectively.
(c,d) Density map of the static susceptibility $\chi_0$ and the relaxation time $\tau$ obtained from the Monte Carlo calculation 
at $J'=0$ (square lattice), the latter from the kinetic TRI protocol. 
}
\label{figs2}
\end{figure*}
\begin{figure*}[tbp]
\begin{center}
\includegraphics[width=17.5cm]{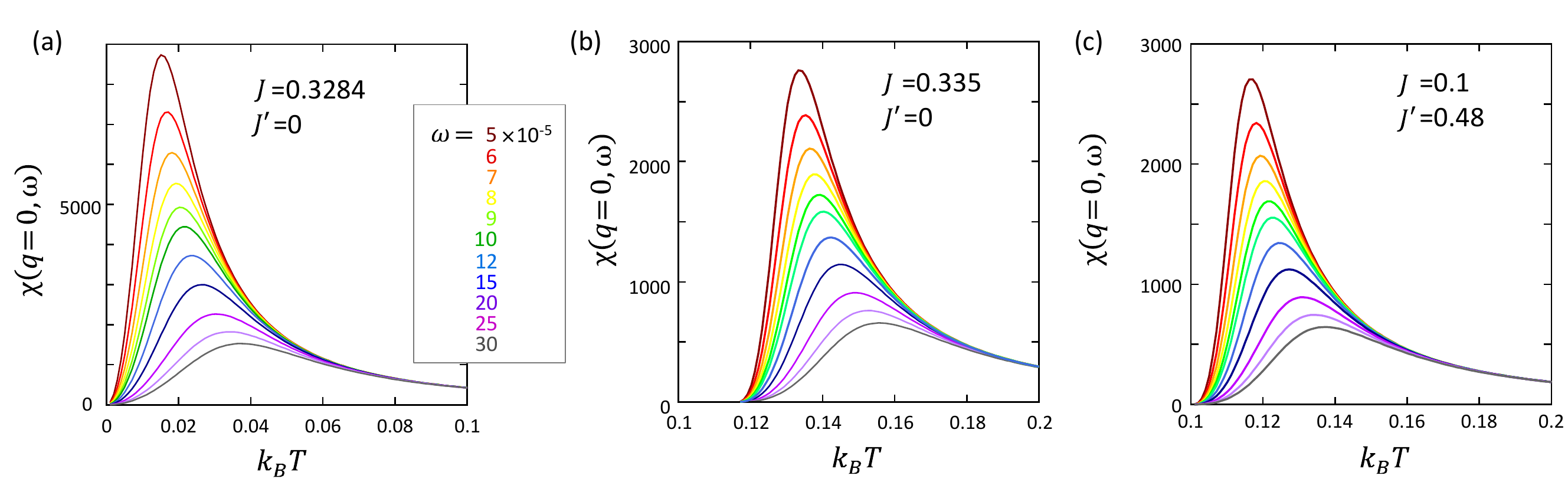}
\end{center}
\caption{
Dynamic susceptibility, $\chi(q,\omega)$, as a function of $k_BT$ for three different choices of $J$ an $J'$ with $\Gamma=1$. 
(a) Square lattice at QCP, $J=0.3284$, $J'=0$, (b) square lattice off QCP, $J=0.3354$, $J'=0$, with $k_BT_c=0.1206$ 
and (c) triangular lattice at QCP, $J=0.1$, $J'=0.48$, with $k_BT_c=0.1163$. 
Panels (b) and (c) follow the 2D Ising universality class. 
}
\label{figs3}
\end{figure*}
\begin{figure*}[tbp]
\begin{center}
\includegraphics[width=17cm]{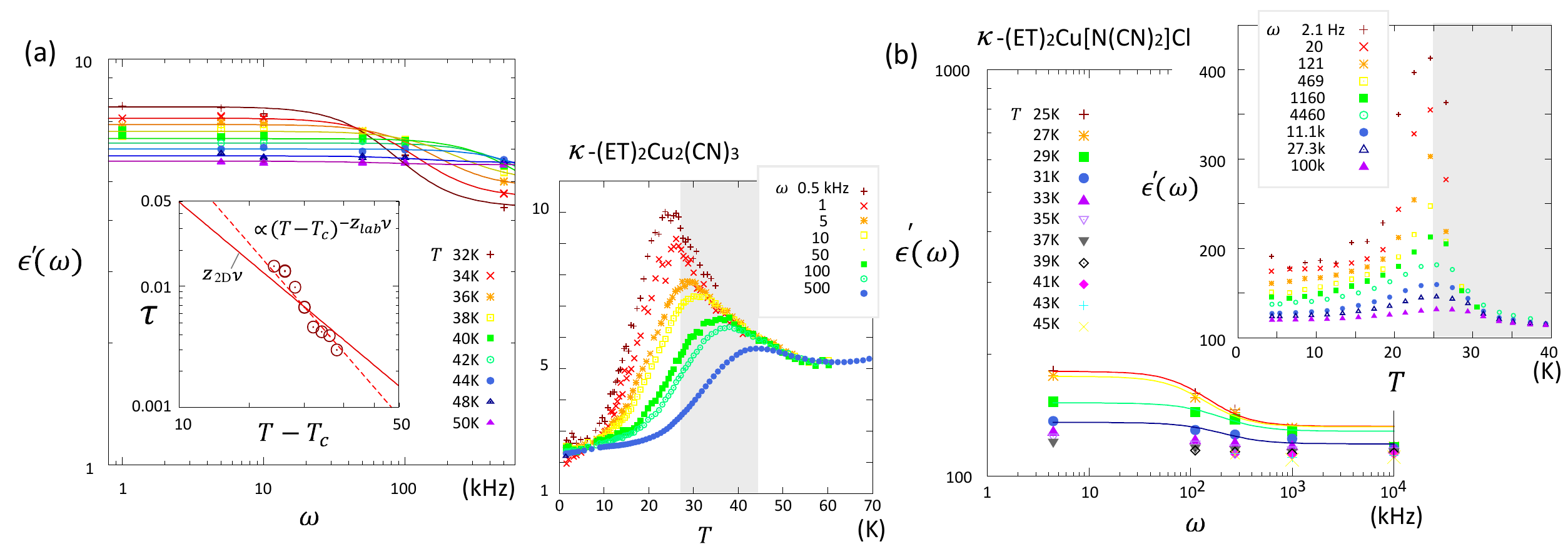}
\end{center}
\caption{Dielectric constant $\epsilon'(\omega)$ in unit of $\epsilon_0$ 
of (a)$\kappa$-ET$_2$ Cu$_2$(CN)$_3$ and (b)$\kappa$-ET$_2$ Cu [N(CN)$_2$]Cl , 
reported in Refs.[\onlinecite{Abdel-Jawad2010},\onlinecite{Lunkenheimer2012}]. 
Data is provided by the courtesy of T. Sasaki and P. Lunkenheimer. 
Solid lines are the Lorentzian fit (Eq.(\ref{eq:5})), 
and the value of $\tau_{lab}$ for (a) is given in the inset as functions of $(T-T_c)$ with $T_c=$6K. 
The solid/broken lines in the inset show the function, $(T-T_c)^{-z\nu}$ with $z=2.18$(2D Ising universality) and 3, respectively. 
}
\label{figs4}
\end{figure*}
\par
Let us consider the strong coupling case, $U, V_1 \gg V_i, t_i$, where 
the electrons do not occupy the same site nor the dimer. 
There are $4^2=16$ basis states in a dimer, but is reduced to four 
in the strong coupling case (see Fig.\ref{figs1}(b)). 
One of the authors has derived the effective Hamiltonian by the perturbation up to the fourth order\cite{Chisa2010}, 
where the second-order perturbation is responsible for the coupling of the spin and 
charge degrees of freedom. 
Whereas, the leading order (namely first order in $t_{ij}$) does not include the spin operator, 
as the spins can only hop within dimers. 
Therefore, taking only the lowest order reduces the number of basis per dimer to two, 
in which the configuration of charge degrees of freedom in the dimer is represented 
via up and down of pseudo-spins, $\sigma_i^z =\pm 1/2$. 
The effective Hamiltonian is reduced to the representation of $m=1\sim 2^N$ basis, 
\begin{equation}
{\mathcal H}_{\rm eff}^{(1)}= \sum_{m,m'} \langle m| H_{mm'} |m'\rangle 
= \sum_{i} -J_{ij} \sigma_i^z \sigma_j^z + \Gamma \sum_{i=1}^N \sigma_i^z
\end{equation}
where $\Gamma=t_1$ and $J=(V_3-V_4)/4$ , $J'=V_2/4$. 
The intra-dimer transfer integral moves the charge back and forth which works as a transverse field that flips the pseudo spins. 
Regarding the inter-dimer interaction, the energy difference between the two different classical configurations of 
pseudo-spins amount to $2J_{ij}$, which are given by that of the original Hamiltonian as 
the difference of contributions from the inter-dimer Coulomb terms. 
As shown in Fig.\ref{figs1}(c), there are four configurations of the adjacent dimers A and B, and only the third panel gives $V_2$ and others zero, 
which is described by the pseudo spin operators as, $V_2(1+\sigma_{A}^z\sigma_{B}^z\;(1-\sigma_{A}^z)/2)/2$. 
As $(\sigma_A^z)^2=1$ and $\langle \sum_{i=1}^N \sigma_i^z \rangle=0$ this term is reduced to  $V_2\sigma_{A}^z\sigma_{B}^z/4$, 
and we find $J'=V_2/4$. The relation, $J=(V_3-V_4)/4$, 
is constructed in the same manner using $V_3$ and $V_4$. 
\par
Substituting the first principles values of $V$'s to the above relation yields, 
$J/\Gamma \sim 0.1$ and $J'/\Gamma \sim 0.5$ for $\Gamma \sim$200 meV, 
and $X=$Cu$_2$(CN)$_3$ has slightly larger values than Cu [N (CN)$_2$]Cl. 
Importantly, it locates in the very vicinity of the QCP ($J/\Gamma=0.1, J'/\Gamma=0.47$) 
in the phase diagram we obtained in Fig.~\ref{f2}(b). 
\par
We briefly note that Ref.[\onlinecite{Kennett2022}] performing a higher order perturbation 
with extra terms included compared to Ref.[\onlinecite{Chisa2010}]. 
Here, we neglect the electron spin degrees of freedom. 

%
%
\section{Square lattice transverse Ising model}
\label{app:3}
We study some other parameters in the phase diagram in Fig.~\ref{f2}(b), 
the square lattice ferromagnetic transverse Ising model at $J'=0$. 
Qualitatively the same results are obtained for the square lattice. 
Figures \ref{figs2}(a) and \ref{figs2}(b) are the $k_BT$ dependences of $\chi_{0;L}$ and $\tau_L$ 
to be compared with Fig.~\ref{f2}(d). 
Here, we show both the case at QCP and just off QCP, which follow the exponents of the 3D and 2D Ising universality classes, respectively. 
The plots of $\tau$ and $\chi_0$ on the plane of $J$ and $k_BT$ are shown for wider temperature range than Fig.~\ref{f4} 
in the main text. 
Although the contour lines are rather different, the overall tendency does not depend on the parameters $J$ and $J'$. 
Also, $\tau$ and $\chi_0$ extracted from the envelope function of Figs.~\ref{figs2}(a) and \ref{figs2}(b) at QCP 
of the square lattice almost coincides with that of the anisotropic triangular lattice including the constant coefficients. 
\par
We here note that the temperature dependences of $\tau$ and $\chi_0$ at $J'>J'_c$, namely when $T_c>0$, are different from 
those of the quantum critical point discussed in Eq.(\ref{eq:4}) in the main text. 
They follow, 
\begin{equation}
\tau \propto (T-T_c)^{-z\nu},\hspace{2mm} \chi_0 \propto (T-T_c)^{-\gamma}
\end{equation}
with $z\sim 2.18$, $\nu=1$, and $\gamma=1.75$ (see the main text), which belong to the 2D Ising universality class. 
When fixing the temperature and approaching the phase boundary by varying the model parameters, $g=J$ or $J'$, 
they follow, 
\begin{equation}
\tau \propto |g-g_c|^{-z\nu},\hspace{2mm} \chi_0\propto |g-g_c|^{-\gamma}
\end{equation}
where $g_c=J_c$ or $J_c'$ are the phase boundaries.

\section{Dynamical susceptibility off QCP}
\label{app:4}
We here show in Figs.\ref{figs3}(b) and \ref{figs3}(c) the dynamical susceptibility, $\chi(q,\omega)$, as a function of $k_BT$ 
when the model parameter is slightly off QCP. 
The one at QCP for the square lattice is given together in Fig.\ref{figs3}(a), which is almost the same as that of 
Fig.~\ref{f3}(c) in the main text. 
In the case off QCP, $\chi_0$ and $\tau$ diverge toward $T_c >0$, and below $T_c$, enter the ferro-ordered phase. 
A similar behavior as that of the QCP is observed, but their critical exponents are that of the 2D universality class, 
which we confirmed in the calculation in Fig.\ref{figs2}. 

%
\section{Reexamination of the experimental results by Abdel-Jawad, {\it et. al} and Lunkenheimer, {\it et. al}}
\label{app:5}
Based on our theoretical findings, we here reexamine the previous reports on the dielectric measurements of 
$\kappa$-ET$_2$ Cu$_2$(CN)$_3$ by Majed, {\it et. al} and $\kappa$-ET$_2$ Cu [N(CN)$_2$]Cl by Lunkenheimer, {\it et. al}. 
In these measurements, the dielectric constants in unit of $\epsilon_0$ shows a peak at temperature, $T_m(\omega)$, 
which distributes at 20-50 K in the former and 25-30K in the latter material, 
when the frequency varies from the order of 1Hz to 100kHz (see the insets of Fig.~\ref{figs4}). 
These results shall be qualitatively compared to our $\chi(q=0,\omega)$ besides the constant and 
the possible experimental background values of $\epsilon$'s from a different origin. 
Let us fix the value of $T$ and extract the experimental data from these figures, 
and by plotting them against $\omega$ we find Figs.\ref{figs4}(a) and \ref{figs4}(b). 
In the case of $\kappa$-ET$_2$ Cu$_2$(CN)$_3$, the successive crossing of lines
belonging to different $T$ takes place over the frequency range of 10-500 kHz 
to be compared with Fig.~\ref{f3}(c), which can be the origin of the large frequency dependence of $T_m$. 
These lines are Lorentzian fit following Eq.(\ref{eq:5}) in the main text, 
and the obtained $\tau_{lab}$ (inset of Fig.\ref{figs4}(a), in unit of (kHz)$^{-1}$) 
varies by one order of magnitude during the temperature change of 10K. 
We plot $\tau_{lab}$ against $(T-T_c)$ with $T_c=6$ K, and draw a line proportional to $(T-T_c)^{-z_{lab}\nu}$ with $\nu=1$. 
While we cannot precisely determine the exponents as we are not able to extract reliable error bars 
in fitting $\epsilon$ with relatively small numbers of data points, 
the data seems to fall between $z_{lab}\sim 2.18$(2D critical exponent, solid line)-3(broken line). 
By contrast, in the case of $\kappa$-ET$_2$ Cu [N(CN)$_2$]Cl, such crossing does not take place, 
and $\tau$ stays extremely small of order-10$^{-6}$ (Hz)$^{-1}$ with no significant variation against $T$. 
\\
We thus consider that $\kappa$-ET$_2$ Cu$_2$(CN)$_3$ in the critical region of the phase diagram, 
and the frequency dependence is overall understood as a signature of the dynamical criticality. 
Whereas, the interpretation of $\kappa$-ET$_2$ Cu [N(CN)$_2$]Cl, is not straightforward. 
The almost frequency-independent behavior indicates that the system is in the 
disordered phase slightly of the critical region, whereas $\tau$ is very large. 
One way to reconcile these two tendencies is to notice that 
$\kappa$-ET$_2$ Cu [N(CN)$_2$]Cl has a N\'eel order at 27K, which may be related to the dielectric ordering. 
If the system is near but off the critical point, the coupling of dipoles with spin degrees of freedom may work as a perturbation 
to drive the system to the first-order transition of dipoles and magnetism. 
These couplings indeed emerge in the model one of the authors discussed previously\cite{Chisa2010,Kennett2022}. 
In fact, $\kappa$-ET$_2$ Cu$_2$(CN)$_3$ does not show magnetic ordering down to lowest temperature, 
which supports this scenario. 
The first principles calculation shows that $\Gamma=t_1$ is slightly larger, namely $J/\Gamma$ is smaller, 
for $\kappa$-ET$_2$ Cu [N(CN)$_2$]Cl than $\kappa$-ET$_2$ Cu$_2$(CN)$_3$. 
This is also consistent with the fact that the former is off the critical point. 

\bibliography{criticality}
\bibliographystyle{apsrev4-1_mk.bst}
\end{document}